\documentclass[]{spie}  

\usepackage{amsmath,amsfonts,amssymb}
\usepackage{graphicx}
\usepackage[colorlinks=true, allcolors=blue]{hyperref}
\usepackage[nolist,nohyperlinks]{acronym}

\usepackage[lofdepth,lotdepth]{subcaption}
\usepackage{multirow}
\usepackage{siunitx}
\usepackage{amsmath}
\newsavebox{\imagebox}

\graphicspath{ {./images/} }

\acrodef{CMB}{\emph{Cosmic Microwave Background}}
\acrodef{TES}{\emph{Transition Edge Sensor}}
\acrodefplural{TES}{\emph{Transition Edge Sensors}}
\acrodef{LEKID}{\emph{Lumped Element Kinetic Inductance Detector}}
\acrodef{ARC}{\emph{Anti-Reflection Coating}}
\acrodef{CPW}{\emph{Co-Planar Waveguide}}
\acrodef{TL}{\emph{Transmission Line}}
\acrodef{CMB}{\emph{Cosmic Microwave Background}}
\acrodef{FEM}{\emph{Finite Element Method}}
\acrodef{HMS}{\emph{Huygens Meta-Surface}}
\acrodef{FSS}{\emph{Frequency Selective Surface}}
\acrodefplural{FSS}{\emph{Frequency Selective Surfaces}}
\acrodef{GRIN}{\emph{GRadient INdex}}
\acrodef{MoM}{\emph{Method of Moments}}
\acrodef{FDTD}{\emph{Finite-Domain Time-Domain}}
\acrodef{HFSS}{\emph{High Frequency Structure Simulator}}
\acrodef{FWHM}{\emph{Full Width at Half Maximum}}
\acrodef{FWE2}{\emph{Full Width at $1/e^2$}}
\acrodef{IDL}{\emph{Interactive Data Language}}
\acrodef{CNC}{\emph{Computer Numerical Control}}
\acrodef{VNA}{\emph{Vector Network Analyser}}
\acrodef{DUT}{\emph{Device Under Test}}
\acrodef{Tx}{\emph{Transmitter}}
\acrodef{Rx}{\emph{Receiver}}
\acrodef{SPT}{\emph{South Pole Telescope}}
\acrodef{ACT}{\emph{Atacama Cosmology Telescope}}
\acrodef{CLASS}{\emph{Cosmology Large Angular Scale Surveyor}}
\acrodef{KID}{\emph{Kinetic Inductance Detector}}
\acrodefplural{KID}{\emph{Kinetic Inductance Detectors}}
\acrodef{JCMT}{\emph{James Clerk Maxwell Telescope}}
\acrodef{MUX}{\emph{Multiplexing}}
\acrodef{OMT}{\emph{Ortho-Mode Transducer}}
\acrodef{MCMC}{\emph{Markov Chain Monte Carlo}}
\acrodef{NA}{\emph{Numerical Aperture}}
\acrodef{ATL}{\emph{Amplitude Taper Efficiency}}
\acrodef{PEL}{\emph{Phase Error Efficiency}}
\acrodef{DUT}{\emph{Device Under Test}}
\acrodef{OA}{\emph{Optical Axis}}
\acrodef{RAM}{\emph{Radiation Absorbent Material}}
\acrodef{TRL}{\emph{Technology Readiness Level}}
\acrodef{SNR}{\emph{Signal-to-Noise Ratio}}
\acrodef{CVD}{\emph{Chemical Vapor Deposition}}
\acrodef{DRIE}{Deep Reactive Ion Etching}

\title{Design and experimental investigation of a planar metamaterial Silicon based lenslet}

\author[a]{Thomas Gascard}
\author[a]{Giampaolo Pisano}
\author[a]{Simon Doyle}
\author[a]{Alexey Shitvov}
\author[b]{Jason Austermann}
\author[b]{James Beall}
\author[b]{Johannes Hubmayr}
\author[c]{Benjamin Raymond}
\author[c]{Nils Halverson}
\author[c]{Gregory Jaehnig}
\author[c]{Christopher M. McKenney}
\author[d]{Aritoki Suzuki}

\affil[a]{Astronomy \& Instrumentation Group, Cardiff University, School of Physics and Astronomy, The Parade, CF24 3AA, Cardiff, United-Kingdom}
\affil[b]{National Institute of Standards and Technology, 325 Broadway, Boulder, CO 80305, USA}
\affil[c]{Center for Astrophysics and Space Astronomy, University of Colorado, Boulder, CO, USA}
\affil[d]{Lawrence Berkeley National Laboratory, Cyclotron Rd, Berkeley, CA 94720, USA}

\authorinfo{Further author information: (Send correspondence to Thomas Gascard)\\ E-mail: thomas.gascard@astro.cf.ac.uk}

\begin{document} 
\maketitle

\begin{abstract}
	The next generations of ground-based cosmic microwave background experiments will require polarisation sensitive, multichroic pixels of large focal planes comprising several thousand detectors operating at the photon noise limit. One approach to achieve this goal is to couple light from the telescope to a polarisation sensitive antenna structure connected to a superconducting diplexer network where the desired frequency bands are filtered before being fed to individual ultra-sensitive detectors such as Transition Edge Sensors. Traditionally, arrays constituted of horn antennas, planar phased antennas or anti-reflection coated micro-lenses have been placed in front of planar antenna structures to achieve the gain required to couple efficiently to the telescope optics. In this paper are presented the design concept and a preliminary analysis of the measured performances of a phase-engineered metamaterial flat-lenslet. The flat lens design is inherently matched to free space, avoiding the necessity of an anti-reflection coating layer. It can be fabricated lithographically, making scaling to large format arrays relatively simple. Furthermore, this technology is compatible with the fabrication process required for the production of large-format lumped element kinetic inductance detector arrays which have already demonstrated the required sensitivity along with multiplexing ratios of order 1000 detectors/channel.
\end{abstract}


\section{INTRODUCTION}
\label{sec:intro} 
Observing the polarisation modes present in the \ac{CMB} is an essential step for modern cosmology and calls for precision instruments for which novel technologies are actively investigated, particularly focal plane optical coupling and detector technologies. Focal optics influences the angular response, polarisation properties, bandwidth and efficiency of the coupled detectors. Ideally, such system provides a polarisation-symmetric beam response across the full operating bandwidth. The planar metamaterial based air-gap phase-delay lenslet reported here, following up on initial investigations\cite{pisano_development_2020} of a variety of lens design approaches, is a step forward towards enabling large detector array optical coupling necessary for future cosmology experiments.
Current investigations are conducted on various focal optic solutions. Corrugated horns made of stacked Silicon platelets\cite{holland_design_2016} offer excellent polarisation properties, important control over the beam width and do not need \ac{ARC}, but they are constrained in bandwidth and can only be either waveguide fed or directly coupled to the detectors. Phased antenna arrays, as is done for SPIDER\cite{gualtieri_spider_2018} and BICEP2/Keck arrays\cite{array_bicep2_2019} are inherently telecentric, limiting the optical coupling between the detectors and telescope optics. Such technology requires multiple ground planes and microstrip crossovers, leading to a challenging fabrication. At last, detectors can be coupled with an antenna and lens system, as has been done for POLARBEAR2\cite{holland_polarbear-2_2016} and \ac{SPT}-3G\cite{carter_year_2018}. This approach allows the antenna to radiate through a dielectric such as Silicon onto which the lens can sit, increasing the gain response proportionally to the index of refraction while allowing for a smaller antenna design, thus providing extra room for detectors, microstrip based filters, and others. However, the hyper-hemispherical lenses require deposition of an \ac{ARC} layer which reduces the bandwidth, the spherical surface is rough due to fabrication limits, degrading the performances. Assembling lenslets into an array induces stray light issues in the detector wafer\cite{yates_surface_2017} if not compensated by an additional resistive layer. The phase engineered planar meta-lenslet reported here provides a valuable workaround, they share the detector plane fabrication process, offering an integrated solution with attractive manufacturing tolerance, good optical performance, low polarisation systematics and an embedded \ac{ARC}. Such device can further be coupled with any desired feed while maintaining broadband capabilities.
The meta-lens \ac{TRL} is at 2, awaiting experimental validation and on-sky demonstration to progress. In this paper will be presented the phase-engineered design principle behind the device along with preliminary measurements of its performances.

\section{METAMATERIAL LENSLET DESIGN}
\label{sec:metalens}
\begin{figure}[h!]
	\centering\includegraphics[width=0.8\textwidth]{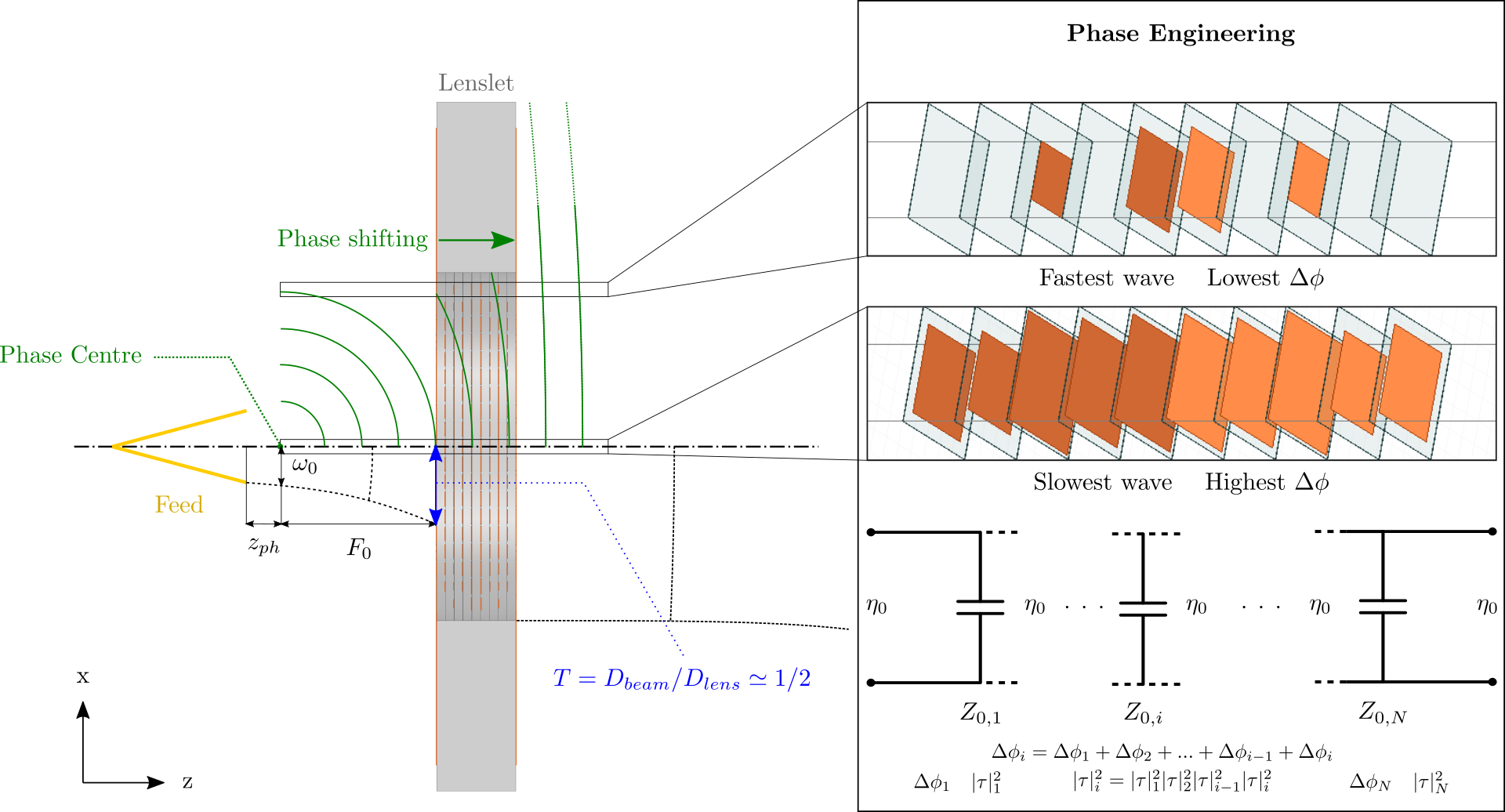}
	\caption{Phase-engineered metamaterial lenslet. The feed provides a gaussian beam with a known phase profile at the distance $Z_0 = z_{ph} + F_0$, determined from the optimal truncation by the lenslet. The phase shifts desired across the lens aperture to obtain a fully transmitted plane wave at the lens output constitutes the validation criteria for the design of each column, realised via a \ac{TL} based code.}
	\label{fig:design1}
\end{figure}

Metamaterials enable striking freedom in shaping light at large scales based on individually tunable minute elements while providing a flat and compact architecture with straightforward manufacturability. When applied to the design of a lens, the approach goes beyond solving conventional lens-maker's equations. The entire structure consist of a three-dimensional array of metal patches arranged in columns, or pixels. Specifically, the lens constitutive elements can be tuned to match any desired feed, including planar antennas, waveguides, etc., as their transversal phase profile is engineered to transform a circular wave to a planar wave, as illustrated in Fig. \ref{fig:design1}. Each column is carefully designed via a \ac{TL} based code derived from the meta-mesh filter design techniques and each pixel may be considered as an isolated phase shifting waveguide. A precursor lenslet fully embedded in polypropylene\cite{pisano_dielectrically_2013}\cite{pisano_planar_2016} has previously been realised using this design approach, demonstrating the working principle presented in this section.

The source at focus initially selected for this lenslet was a chocked circular horn operating in the W-band $f \in [75\si{GHz}; 110\si{GHz}]$. The beam response is determined via \ac{FEM} in \ac{HFSS} \cite{noauthor_ansys_nodate} and is presented in section \ref{sec:lensmeas}, Fig. \ref{fig:lensmeas1}. The largest \ac{FWHM} angle $\vartheta_{FWHM} = 32.5^{\circ}$ can then be extracted and the equivalent beam waist produced by the horn $\omega_0 = 980\si{um}$ is deduced as a function of frequency as per the left part of Eq.\ref{eq:11}\cite{goldsmith_quasioptical_1998}. The Rayleigh distance $z_R^{90\si{GHz}} = 905\si{um}$ is then deduced using the right part of Eq.\ref{eq:11}. It effectively defines the maximal admissible variation for the lens focus position from the probe aperture along the optical axis. Additionally, the phase centre of this probe, origin of the gaussian beam, is obtained from \ac{HFSS} simulation as the peak-to-peak phase difference of the E-field, found at $z-{ph} = 1.697\si{mm}$ in front of its aperture.
\begin{equation}
	\omega_0(\lambda) = \frac{\lambda\sqrt{2\ln(2)}}{2\pi\tan(\vartheta_{HWHM})}  \qquad
	z_R = \frac{\pi\omega_0^2(\lambda)}{\lambda}
	\label{eq:11}
\end{equation}

\begin{figure}[h!]
	\centering\includegraphics[width=0.6\textwidth]{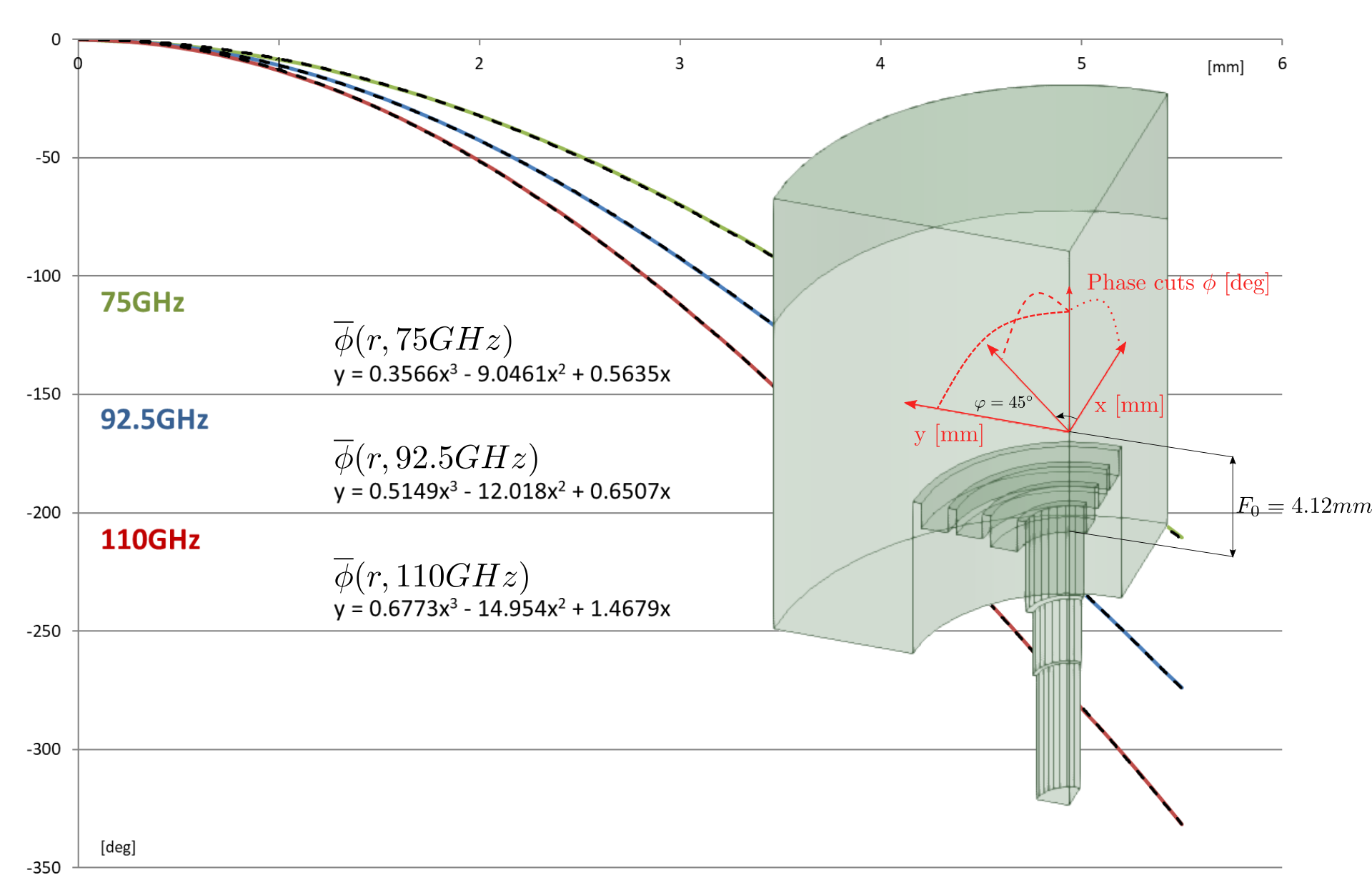}
	\caption{Phase patterns of the feed at the distance $Z_0$ and at a given frequency $f$ along the lenslet radius $r$ for an azimuth angle of $\varphi = 0^{\circ}, 45^{\circ}, 90^{\circ}$. These cuts are averaged together and this average is then subtracted from its central value at the azimuth angle $\theta = 0^{\circ}$. This is then fitted to a polynomial.}
	\label{fig:design2}
\end{figure}

In order to define the lens focal length $F_0$ given its diametre $D = 10mm$ arbitrarily chosen, a first approximation is to consider it as a thin lens truncating the gaussian beam emitted by the probe. As is described in \cite{urey_spot_2004}, such a truncated beam profile stays gaussian for a truncation ratio $T \in [0.4; 0.9]$ and in this interval there is a trade-off between the power loss at the aperture, increasing with $T$, and the peak axial irradiance, decreasing with $T$, with a maximum reached for $T \rightarrow 1$. With this constraint in mind, the optimal focal length can be determined at a given frequency, here taken at 90GHz, using Eq. \ref{eq:1}. 
\begin{equation}
	\omega_0(\lambda) = \frac{\lambda\sqrt{2\ln(2)}}{2\pi\tan(\vartheta_{HWHM})} \quad
	\theta = \frac{\lambda}{\pi\omega_0} \quad
	NA = \sin \theta \quad
	s = 2\omega_0 \quad
	K = \frac{4\omega_0}{\lambda NA} \quad
	\omega(F_0) = \frac{D T}{2} = \omega_0\sqrt{1+\left(\frac{\lambda F_0}{\pi \omega_0^2}\right)^2}
	\label{eq:1}
\end{equation}
The apex half-angle of the probe $\theta = 62^\circ$ gives a \ac{NA} for the lenslet at $NA = 0.88$. Knowing the spot size focused by the lens at the probe aperture should be $s = 1.96\si{mm}$, the related spot-size constant $K = 1.336$ can be found. The latter relation is valid for $f\# < 2$. This constant is a function of beam truncation $T = 2\omega/D$ only, and can be evaluated based on the \ac{FWHM} intensity as $K = 0.75/T$ or on the \ac{FWE2} intensity as $K = 1.27/T$. In practice while maximum peak axial illumination is reached with the latter expression, the first provides optimal trade-off between illumination and power, giving $T = 0.56$. $\omega$ is the beam waist at truncation, positioned at the focal length found to be $F_0 = 2.42\si{mm}$ and giving $f\# = 0.24$. Accounting for an offset centre of phase position, found at $z_{ph} = 1.697\si{mm}$, the optimal distance from the lens to the feed's aperture $Z_0 = z_{ph} + F_0 = 4.12\si{mm}$ is now fully defined. The phase-engineering procedure is driven by the phase response of the source $\phi$ at $Z_0$ on the plane where the lens intersects its beam, obtained in \ac{HFSS}, as presented in Fig. \ref{fig:design2}. A column of patches at a given position along the lens radius $r$ will in fact provide a phase shift $\overline{\Delta \phi}$ derived from a third degree polynomial fit, with coefficients $c_i$, of the phase shift $\Delta \phi = \phi - \phi_{\theta=0^\circ}$ averaged from the response at azimuths $\theta = 0^\circ, 45^\circ$ and $90^\circ$ as per Eq. \ref{eq:2}.
\begin{equation}
	\overline{\Delta \phi}(r,f) = c_3(f)r^3+c_2(f)r^2+c_1(f)r = (\Delta \phi(r,f)_{\theta = 0^\circ} + \Delta \phi(r,f)_{\theta = 45^\circ} + \Delta \phi(r,f)_{\theta =9 0^\circ})/3
	\label{eq:2}
\end{equation}

The code at the core of the phase engineering approach finds its roots in meta-mesh filter design\cite{ulrich_far-infrared_1967}\cite{marcuvitz_waveguide_1986} where a stack of $N$ \acp{FSS} is modelled by a lossless \ac{TL} of equivalent distributed elements. This is effectively representing the wave propagation in the medium between the adjacent grids, connecting lumped shunt admittances emulating electromagnetic response of the patches. Each constituent of the stack can be seen as a low-pass filter matching a given impedance, here $\eta_0$, of resonant angular frequency $\kappa_{0,i} = 1-0.27(a_i/g)$, normalised frequency $\kappa = g/\lambda$, thus a generalised frequency of $\Omega_i = \kappa/\kappa_{0,i} - \kappa_{0,i}/\kappa$. It provides a transmitted phase shift $\Delta \phi_i$, has a characteristic impedance $Z_{0,i}$ and a transmittance $|\tau|^2_i$, functions of the \acp{FSS} geometrical parameters ratio $a_i/g$ with $a_i$ the half-gap between square patches and $g$ is the unit-cell size, as per Eq. \ref{eq:3}. The transmittance is desired above 99\%, requiring $\kappa$ at $0.15$ thus $g = 500\si{um}$. For the column under investigation, each constituent's geometry is refined via a \ac{MCMC} optimisation. To lighten the computing load, E,H,D and central transverse plane symmetries are introduced, bringing the effort down to a half-section of an octant of the lens with $N/2$ elements per columns, where $N$ is now necessarily even. To initiate the code, the optimal number of layers is determined by refining the $a_i/g$ and $N$ of the central column, for which the highest phase shift $2\sum_{i=1}^{N/2} \Delta \phi_i \rightarrow max(|\overline{\Delta \phi}|)$ is desired.
\begin{equation}
	Z_{0,i} = \left[2\ln\left(\csc\left(\frac{\pi}{2}\frac{a_i}{g}\right)\right)\right]^{-1} \qquad
	\Delta \phi_i = \arctan\left(\frac{1}{Z_{0,i}\Omega_i}\right) \qquad
	|\tau|^2_i = \frac{Z_{0,i}^2\Omega_i^2}{1+Z_{0,i}^2\Omega_i^2} 
	\label{eq:3}
\end{equation}

\begin{figure}[h!]
	\centering
	\savebox{\imagebox}{\includegraphics[width=0.3\textwidth]{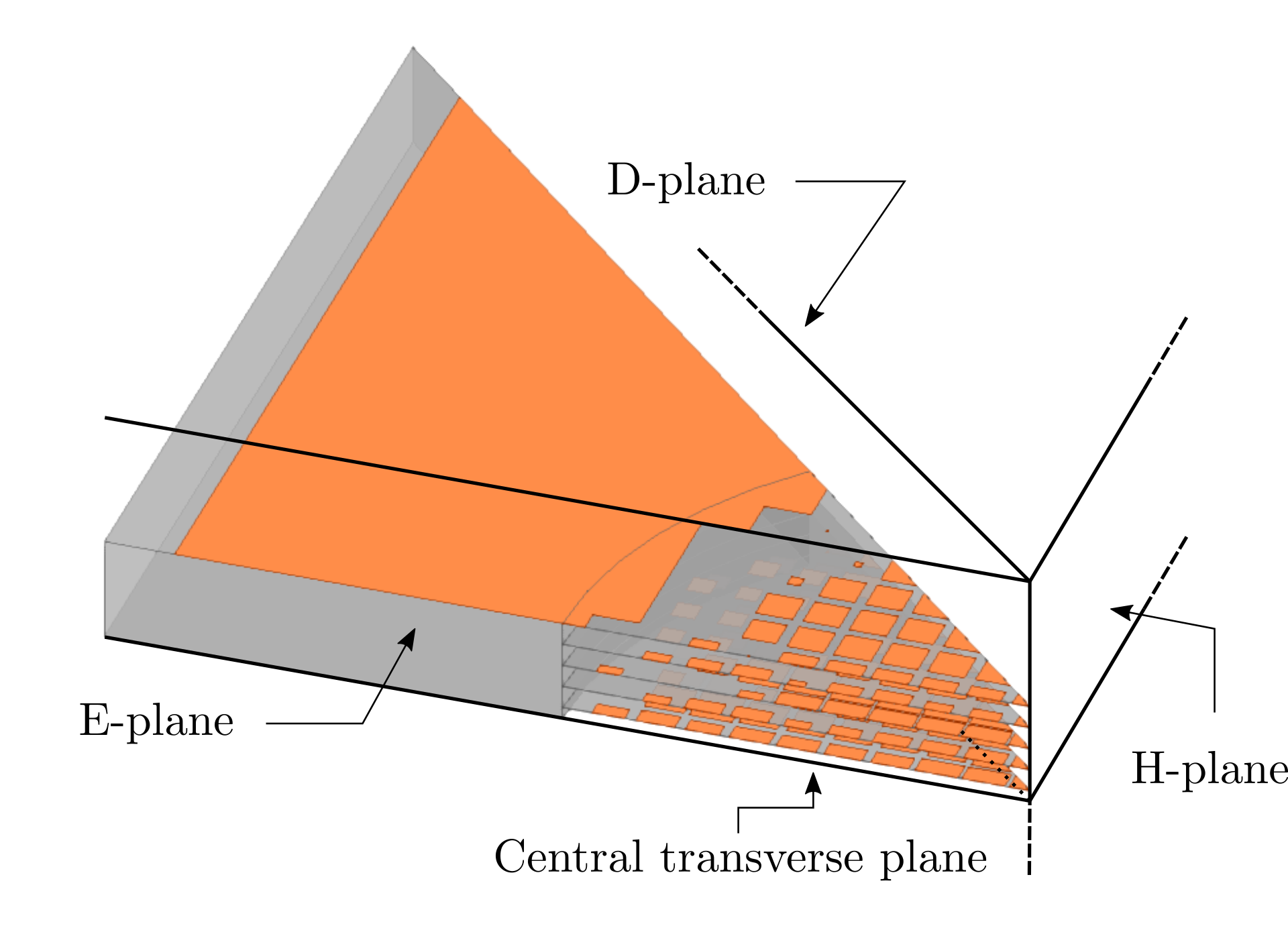}}
	\begin{subfigure}[t]{0.3\textwidth}
		\centering\usebox{\imagebox}
		\caption{}
		\label{fig:design31}
	\end{subfigure}
	\begin{subfigure}[t]{0.4\textwidth}
		\centering\raisebox{\dimexpr.5\ht\imagebox-.5\height}
			{\includegraphics[width=\textwidth]{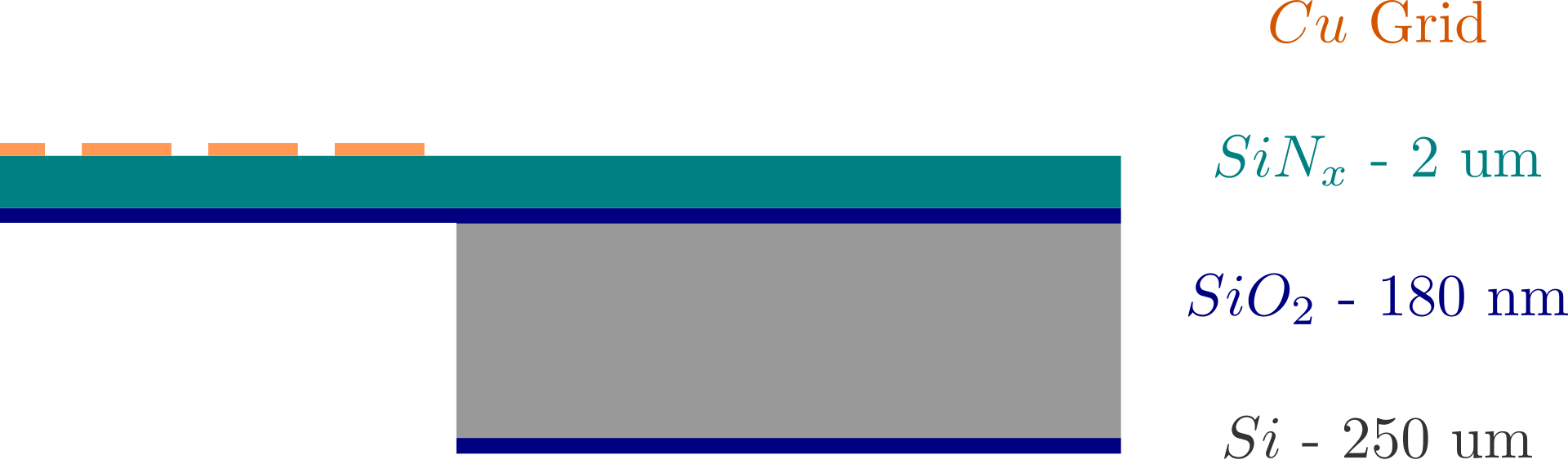}}
		\caption{}
		\label{fig:design32}
	\end{subfigure}
	\begin{subfigure}[t]{0.2\textwidth}
		\centering\raisebox{\dimexpr.5\ht\imagebox-.5\height}
			{\includegraphics[width=\textwidth]{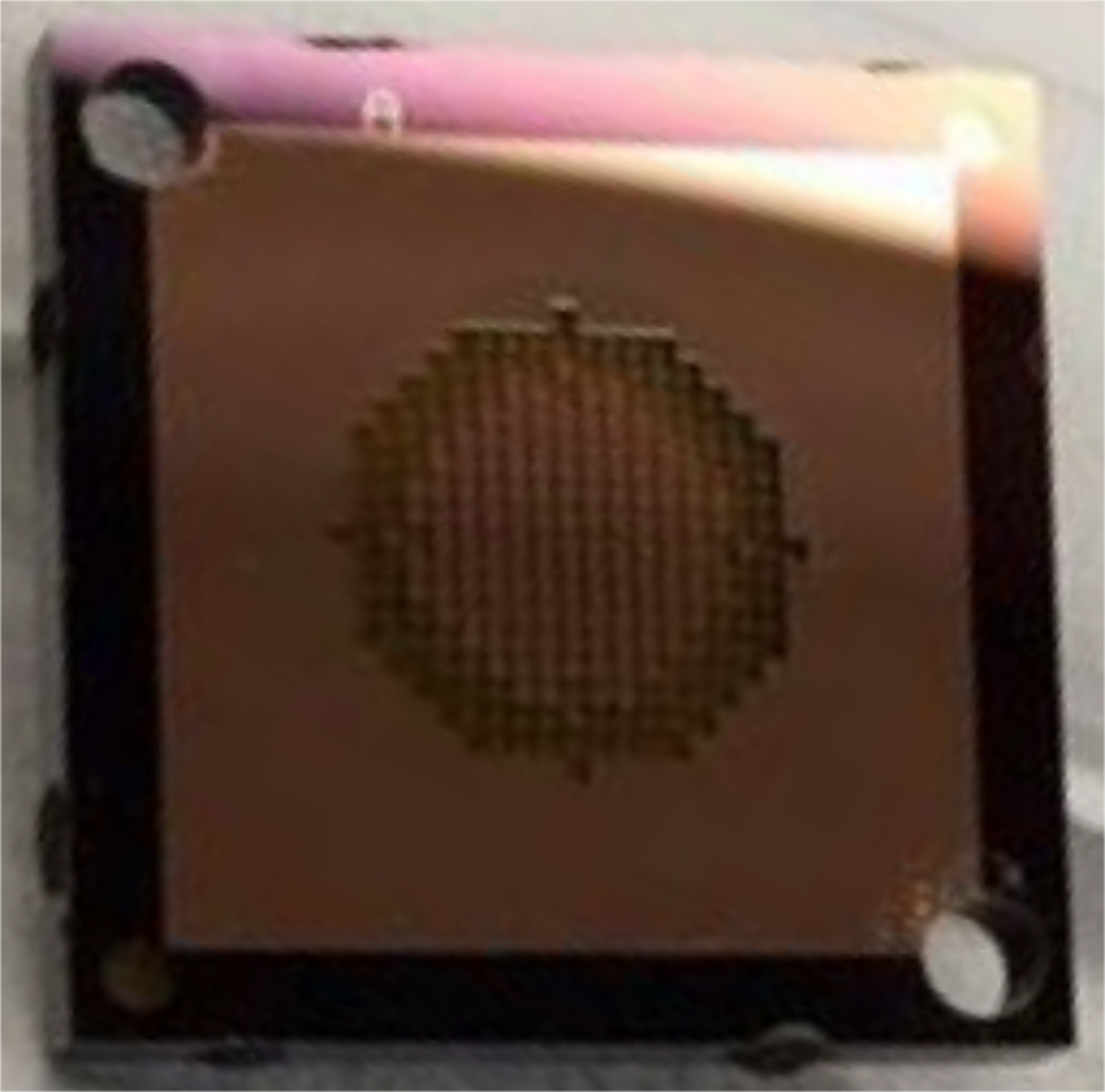}}
		\caption{}
		\label{fig:design33}
	\end{subfigure}
	\caption{\subref{fig:design31} Effective half-octant part of the lens refined and associated symmetry planes. \subref{fig:design32} Schematic cross section of the lenslet with constitutive layers. \subref{fig:design33} Photograph of the fabricated lenslet.}
	\label{fig:design3}
\end{figure}

The half-octant is presented in Fig. \ref{fig:design3} along with the fabricated lens comporting $N = 10$ layers for 40 refined columns. The copper deposition provides a feature thickness at $2\si{um}$. As such material has a skin depth of $0.2\si{um}$ at $100\si{GHz}$, the grids can be considered as infinitely thin. These \acp{FSS} are photolithographed on a $2\si{um}$ thin \ac{CVD} silicon nitride substrate sitting on $250\si{um}$ \ac{DRIE} silicon rims, giving a $250\si{um}$ air-gap between each layer. This lenslet is effectively a demonstrator for the phase-engineering design, it has been compared with the \ac{GRIN} approach \cite{pisano_development_2020}. The next stage will be to generate a structure fully embedded in Silicon, allowing efficient coupling with photolithographed feed such as planar antennas, offering excellent unidirectional radiation in such thick dielectric substrate and flexibility in the underlying microstrip line feeding the detectors.

\section{LENSLET PRELIMINARY MEASUREMENTS}
\label{sec:lensmeas}
Validation of the expected performances of the phase-engineered design approach is an important milestone in the \ac{TRL} progression of the meta-mesh lens. Performances shown in simulations needs to be demonstrated experimentally to reach the third stage. In this section, the simulated device characteristics are presented alongside preliminary experimental validation.

\subsection{Probe characterisation} \label{subsec:probe}
\begin{figure}[h!]
	\centering
	\savebox{\imagebox}{\includegraphics[width=0.5\textwidth]{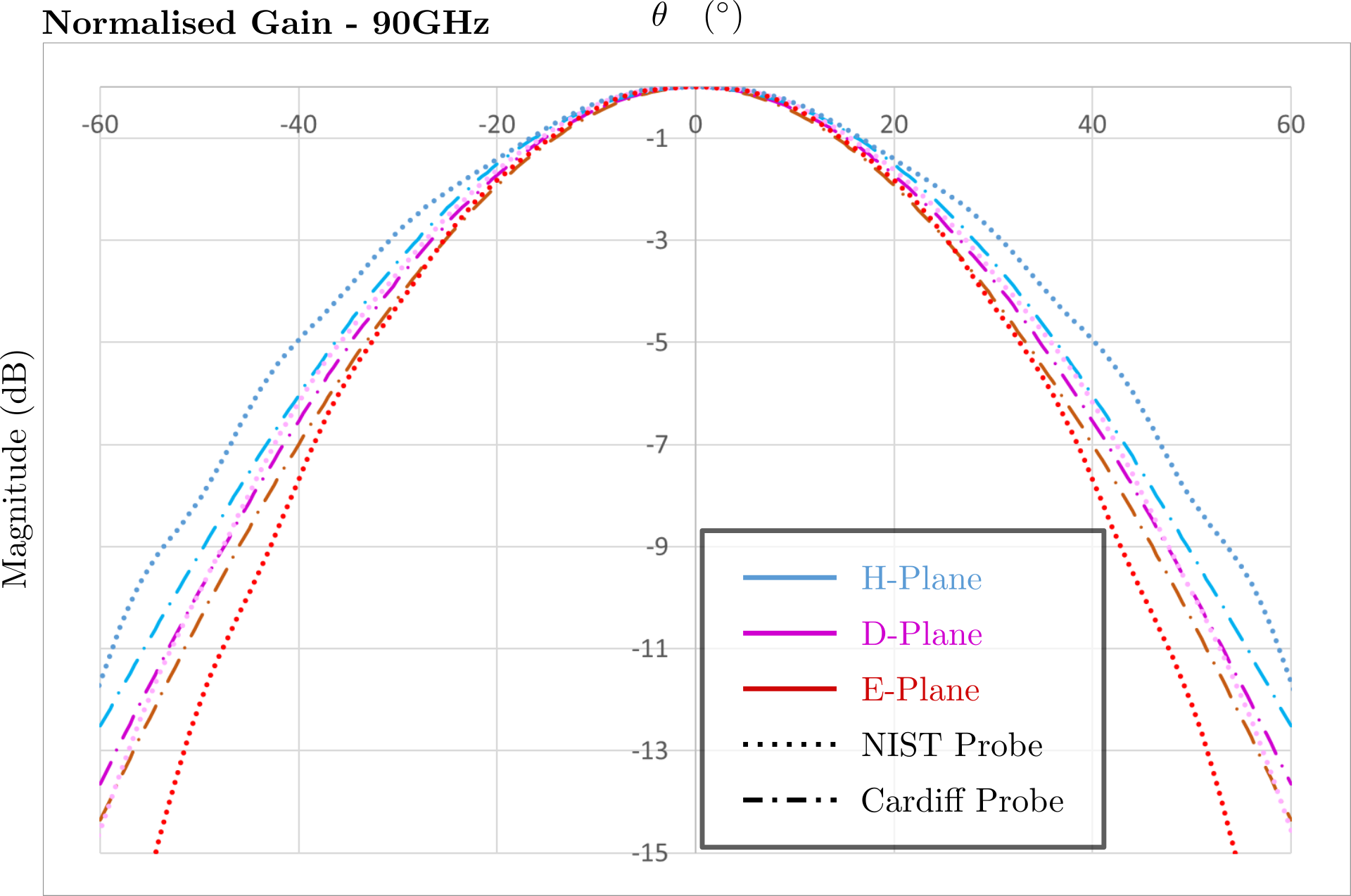}}
	\begin{subfigure}[t]{0.2\textwidth}
		\centering\raisebox{\dimexpr.5\ht\imagebox-.5\height}
		{\includegraphics[width=\textwidth]{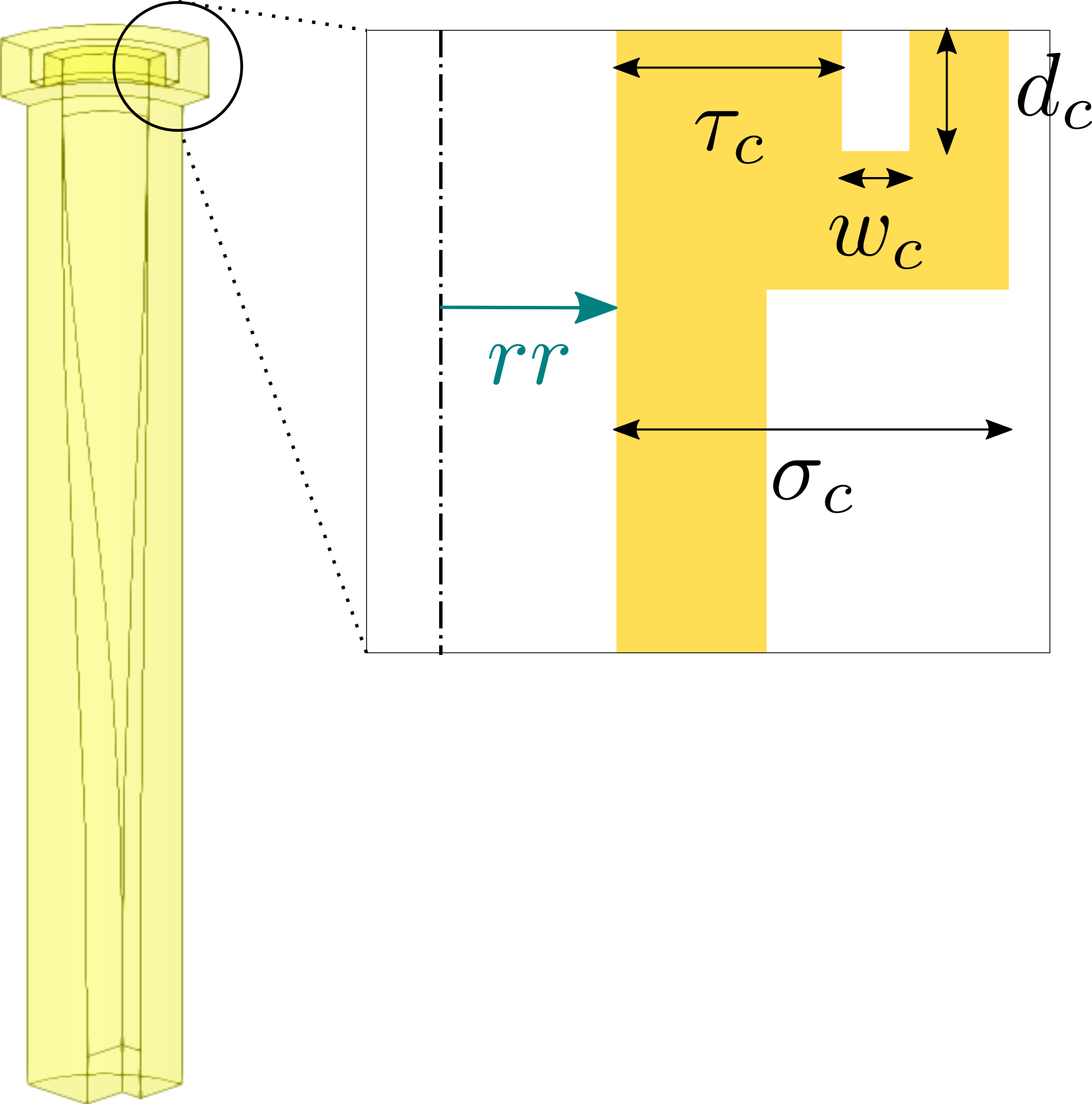}}
		\caption{}
		\label{fig:probe11}
	\end{subfigure}\qquad
	\begin{subfigure}[t]{0.5\textwidth}
		\centering\usebox{\imagebox}
		\caption{}
		\label{fig:probe12}
	\end{subfigure}
	\caption{\subref{fig:probe11} Rectangular to circular transition of input WR10 rectangular waveguide and output circular aperture of radius $rr = 1.55\si{mm}$. The insert underlines the chocke with flange width $\sigma_c$, distance from the waveguide wall $\tau_c$, width $w_c$ and depth $d_c$. \subref{fig:probe12} Compared beam responses of the initial feedhorn (dotted) with the choked waveguide converter at $90\si{GHz}$.}
	\label{fig:lensmeas1}
\end{figure}

\begin{figure}[h!]
	\centering
	\savebox{\imagebox}{\includegraphics[width=0.45\textwidth]{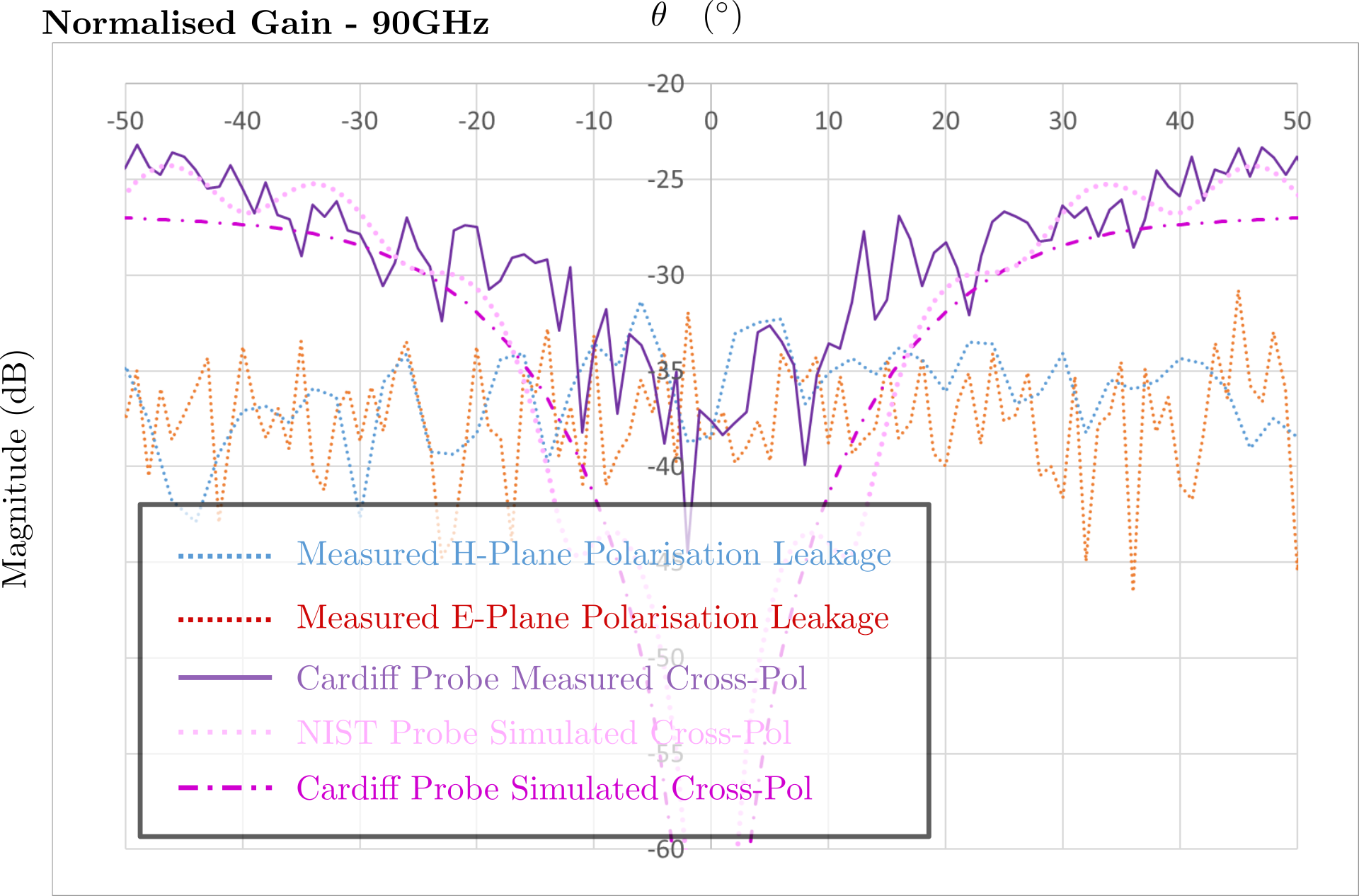}}
	\begin{subfigure}[t]{0.45\textwidth}
		\centering\raisebox{\dimexpr.5\ht\imagebox-.5\height}
		{\includegraphics[width=\textwidth]{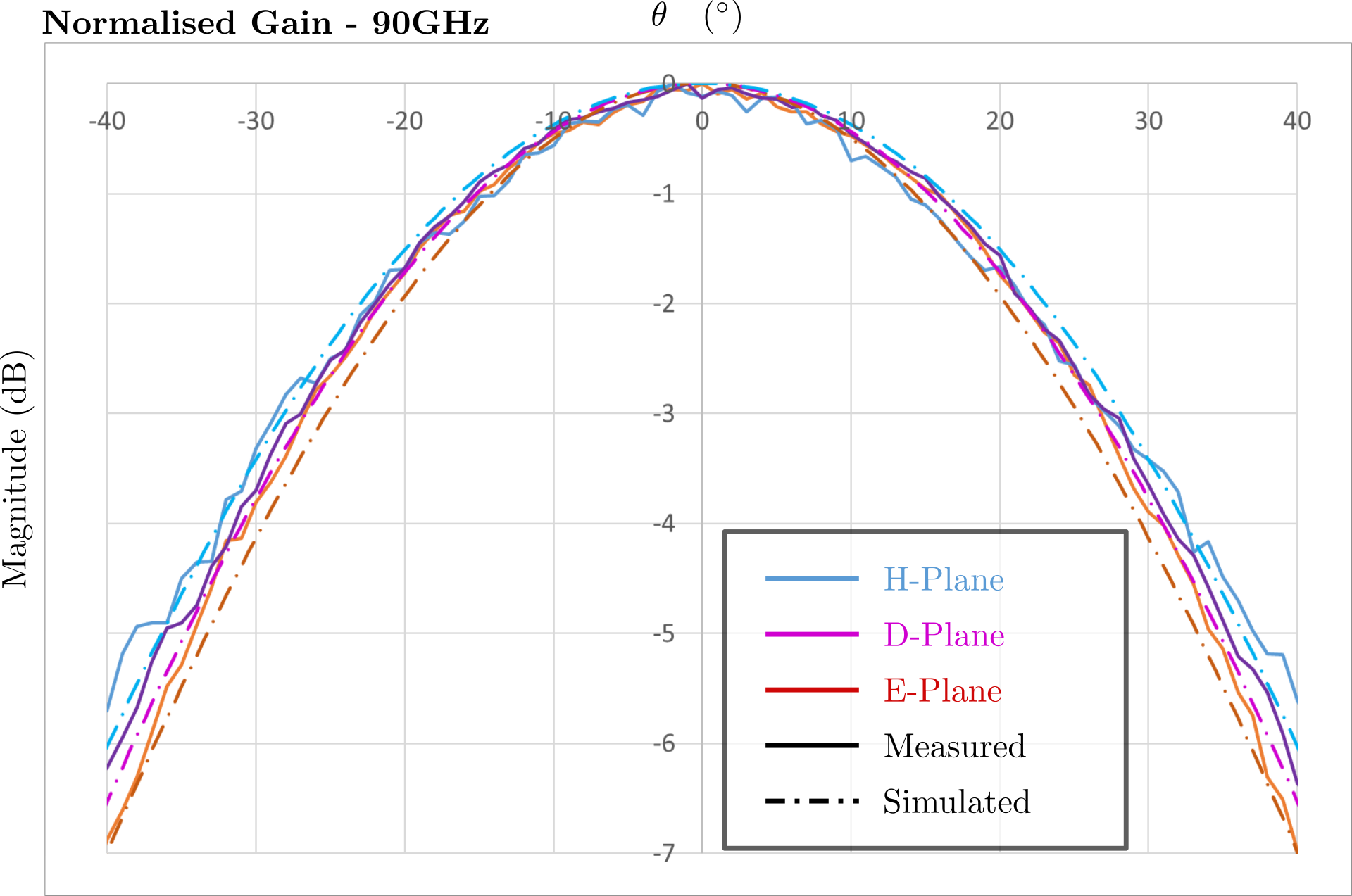}}
		\caption{}
		\label{fig:probem1}
	\end{subfigure}\qquad
	\begin{subfigure}[t]{0.45\textwidth}
		\centering\usebox{\imagebox}
		\caption{}
		\label{fig:probem2}
	\end{subfigure}
	\caption{\subref{fig:probem1} Measured co-polarisation cuts of the waveguide transition in the E, D and H-planes at $90GHz$. \subref{fig:probem2} Measured cross-polarisation (D-plane) and leakage (E and H-planes) cuts of the waveguide transition at $90GHz$ along with the commercial horn cross-polarisation cut for comparison.}
	\label{fig:probemeas1}
\end{figure}

The feeding element coupled to the lenslet is critical, as has been seen in Section \ref{sec:metalens}. Regrettably, the dedicated commercial horn was unavailable in the laboratory. A custom waveguide transition was used instead, transferring light from a WR10 rectangular aperture to a circular aperture of radius $rr = 1.55\si{mm}$, further chocked\cite{james_radiation_1977} to obtain a better homogeneity of the beam response and to reduce cross-polarisation. The choke optimal geometry is given for a flange width $\sigma_c = 2\lambda/3 = 2.2\si{mm}$, choke distance from the waveguide wall $\tau_c = 0.18\lambda = 605\si{um}$, width $w_c = 0.15\lambda = 500\si{um}$ and depth $d_c = 0.265\lambda = 883\si{um}$. The waveguide transition design and simulated beam response are shown in Fig. \ref{fig:lensmeas1} at the frequency $f = 90\si{GHz}$. The patterns can be compared with the aforementioned commercial horn response also present in this figure, showing a similar gaussian beam of waist $\omega_0 = 1.29\si{mm}$. Comparison of the measured and simulated responses of the waveguide transition are provided in Fig. \ref{fig:probemeas1}, showing a cross-polarisation level of $-25\si{dB}$ at worst. Its phase centre position is at the aperture $z_{ph} = 0mm$. Such substitute feed is perfectly suitable to feed the lenslet if positioned at the proper distance found to be at the new focal length $Z_0 = F_0 = 2.3\si{mm}$ derived from Eq. \ref{eq:1}. Measured response of this converter is also showed in Fig. \ref{fig:lensmeas1}, showing excellent agreement for the azimuth range $\theta \in [-50^\circ; 50^\circ]$, with a gaussian fit matched down to $2.6^\circ$ accuracy.

\subsection{Pre-analysis via simulations}  \label{subsec:preanalysis}
\begin{figure}[h!] 
	\centering
	\savebox{\imagebox}{\includegraphics[width=0.7\textwidth]{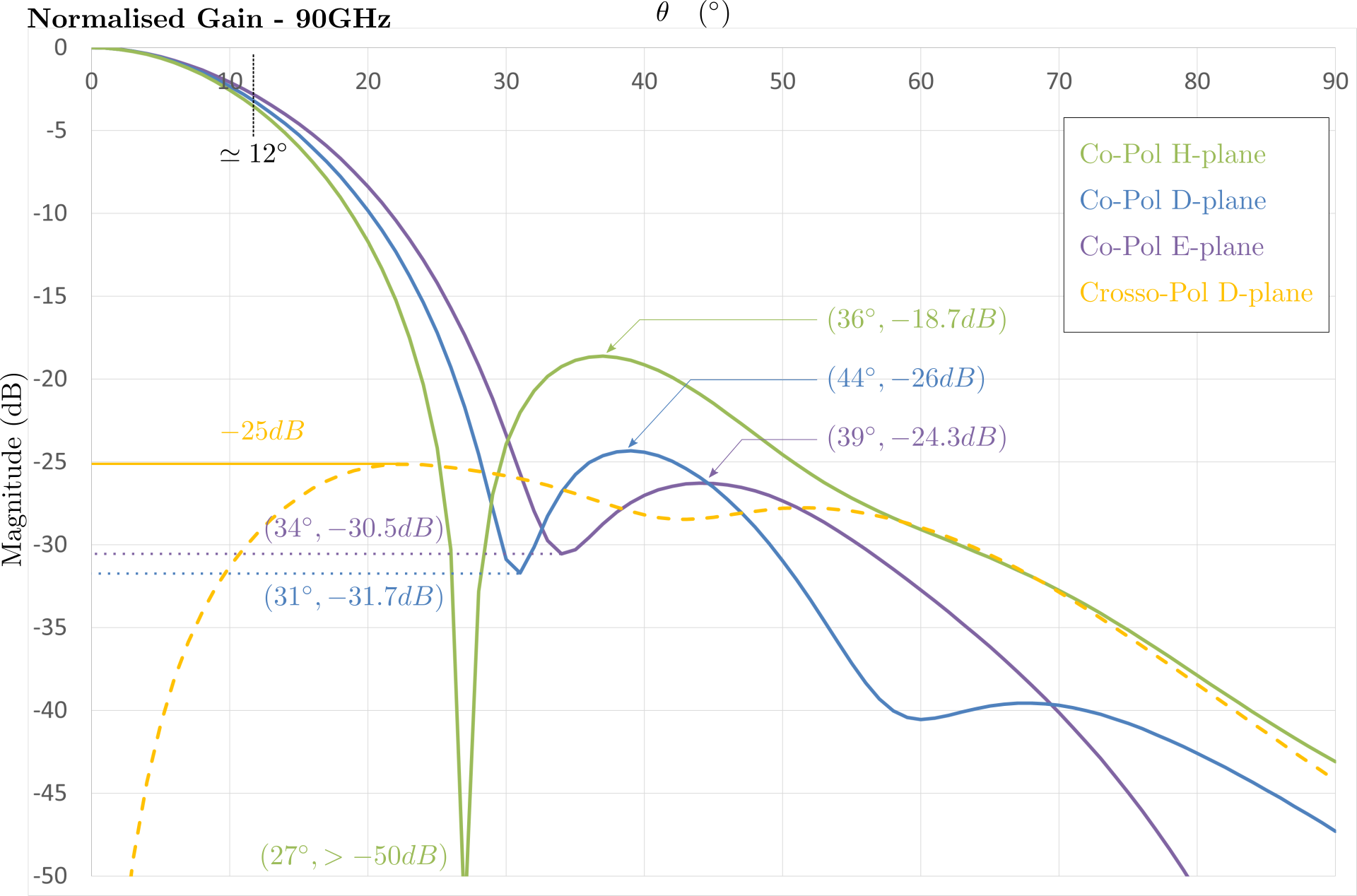}}
	\begin{subfigure}[t]{0.15\textwidth}
		\centering\raisebox{\dimexpr.5\ht\imagebox-.5\height}
		{\includegraphics[width=\textwidth]{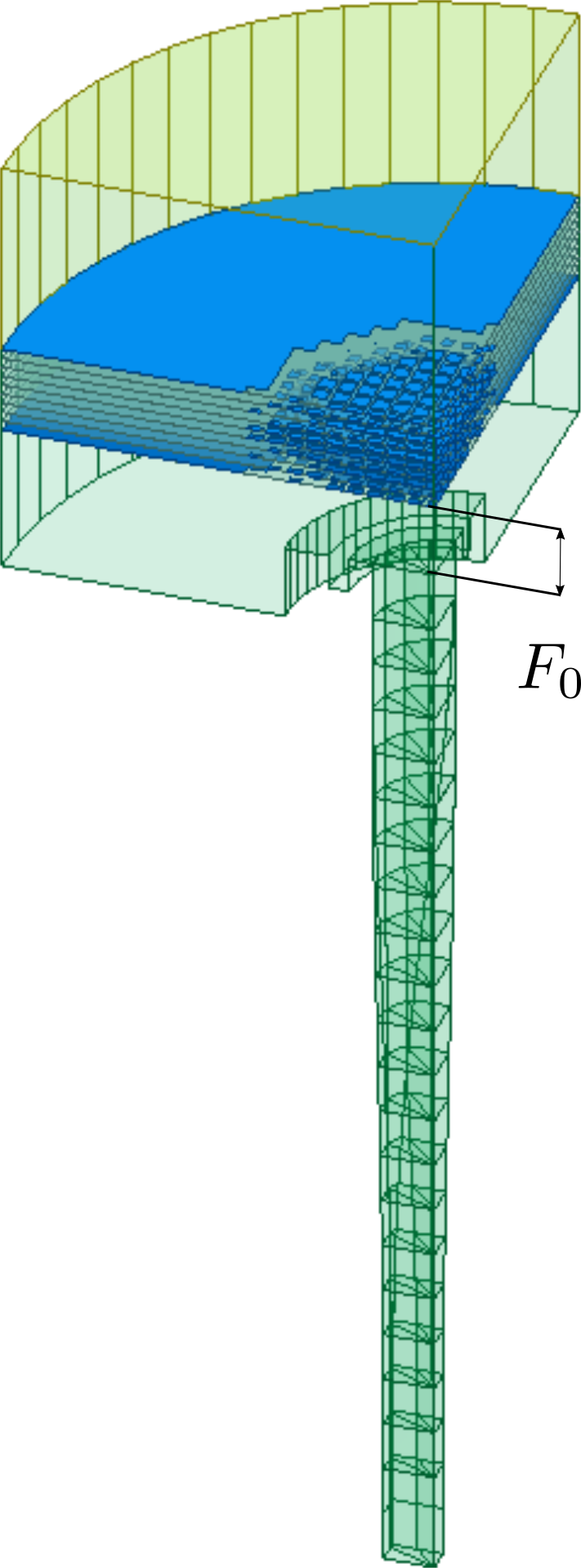}}
		\caption{}
		\label{fig:lensmeas21}
	\end{subfigure}
	\quad
	\begin{subfigure}[t]{0.7\textwidth}
		\centering\usebox{\imagebox}
		\caption{}
		\label{fig:lensmeas22}
	\end{subfigure}
	\caption{\subref{fig:lensmeas21} Model of the lenslet coupled to the waveguide transition. The distance between the bottom side of the lenslet and the aperture of the feed is refined until the optimal $F_0$ is found. \subref{fig:lensmeas22} Beam patterns of the coupled system at $90\si{GHz}$ with the co-polarisation response in the E, H and D-planes and the cross-polarisation.}
	\label{fig:lensmeas2}
\end{figure}

With confidence in the modelled probe feeding the system, coupling with the lenslet can be conducted. A first model is investigated to validate the focal length by refining the separation between the lenslet and the feed's aperture. Once the focal length is confirmed, a fine simulation of the response is carried out and serves as the validation baseline for the measurements conducted in the laboratory. The model is illustrated in Fig. \ref{fig:lensmeas2} and provided an optimal beam response at the expected focal length $F_0 = 2.3\si{mm}$. The selection is made on the maximum peak gain and \ac{FWHM}, the best beam homogeneity comparing the E, H and D-plane responses, the highest beam gaussianity, but also on the lowest levels of cross-polarisation and first sidelobe. The gaussianity of the beam is established to the $n^th$ order by operating a regression analysis using least squares to fit an $n$-modded gaussian function to the actual response and estimate the error. 

The simulated response is as expected for a lens, with $f\# = 0.23$, providing good gaussianity and homogeneity between the different planes down to $-30\si{dB}$, with a peak gain at $17.5\si{dB}$ or $8.75\si{dBi}$, first null levels are below $-30\si{dB}$ and reached around $\theta = 30^\circ$. The \ac{FWHM} is $\theta_{FWHM} = 24^\circ$, first sidelobe levels are equal to or below $-18.7\si{dB}$, and the cross-polarisation level is $-25\si{dB}$ at worst. This analytical validation of the phase-engineered lenslet design principle can now be compared with measured data.

\subsection{Procedure} \label{subsec:procedure}
\begin{figure}[h!]
	\centering\includegraphics[width=0.6\textwidth]{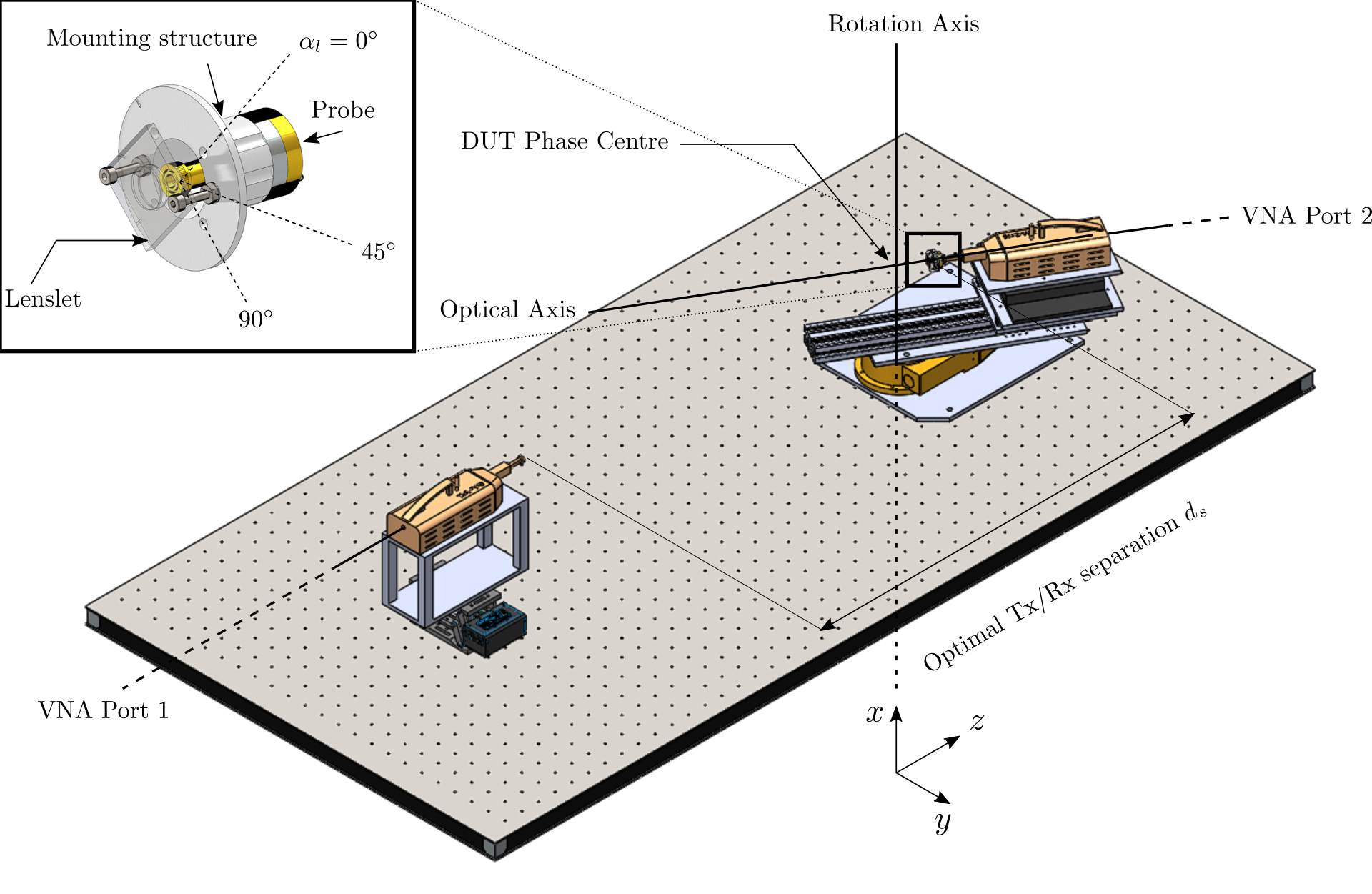}
	\caption{Experimental setup implementing a 2D beam scan through the elevation interval $[-50^\circ; 50^\circ]$ in the H-plane. The beam-cut is selected by rotating the heads at the desired angle with respect to the \ac{OA}.}
	\label{fig:expsub1}
\end{figure}

In order to map the co and cross-polarisation beams on the E, H and D-planes while covering the W-band $f\in[75\si{GHz};110\si{GHz}]$, the experimental setup illustrated in Fig. \ref{fig:expsub1} is designed so that a stationary \ac{Tx} feeds a divergent beam to the \ac{DUT}. The outgoing plane wave is transmitted to a \ac{Rx} positioned in the device's far field. The lenslet and the \ac{Tx} are assembled together as shown in the insert on Fig. \ref{fig:expsub1} and mounted on a rotary stage, enabling the assembly to rotate through the desired elevation range around the device's phase centre on the $zy$-plane. The \ac{OA} is then defined by the horizontal $z$ axis going through the \ac{DUT}, \ac{Tx} and \ac{Rx} phase centres when aligned together. The lenslet can in principle be rotated around the \ac{OA} at an angle $\alpha_l = 0^\circ$, $45^\circ$ or $90^\circ$. To cover the W-band a set of Rhode \& Schwarz (R\&S) \cite{international_accueilgroup_nodate} ZVA Z110 frequency converters are used along with the R\&S ZVA67 \ac{VNA}. Optimal standing waves attenuation versus dynamic range is achieved at a certain distance between the \ac{Rx} and \ac{Tx} probes, here found at $d_s = 50\si{cm}$. Alignment of the mechanical parts on all three axis is done using a laser leveller and fine positioning of the combined \ac{DUT} and \ac{Tx} assembly is conducted so that the \ac{DUT} phase centre is on the axis of rotation using a micrometric $xy$ linear stage. Noise reduction is done using \ac{RAM} extensively to cover reflective area contributing to the device far field and the feed's gaussian beam. Measurements are taken via a dedicated software enabling the automated rotation of the \ac{DUT} and \ac{Rx} while compiling the transmission from port 1 to port 2 $S21$ taken from the \ac{VNA}. The proprietary software provides dataset of 1001 frequency points over the full bandwidth taken back and forth through the elevation interval $[-50^\circ; 50^\circ]$ two times, for the left-hand and right-hand rotations. An averaging over these four passes is done in the post-processing, aiding systematic noise reduction to the VNA embedded averaging capability, set at 16, which gives a total of 64 averages on the data.

\subsection{Measurement results analysis} \label{subsec:results}
\begin{figure}[h!]
	\centering
	\begin{subfigure}[t]{0.45\textwidth}
		\includegraphics[width=\textwidth]{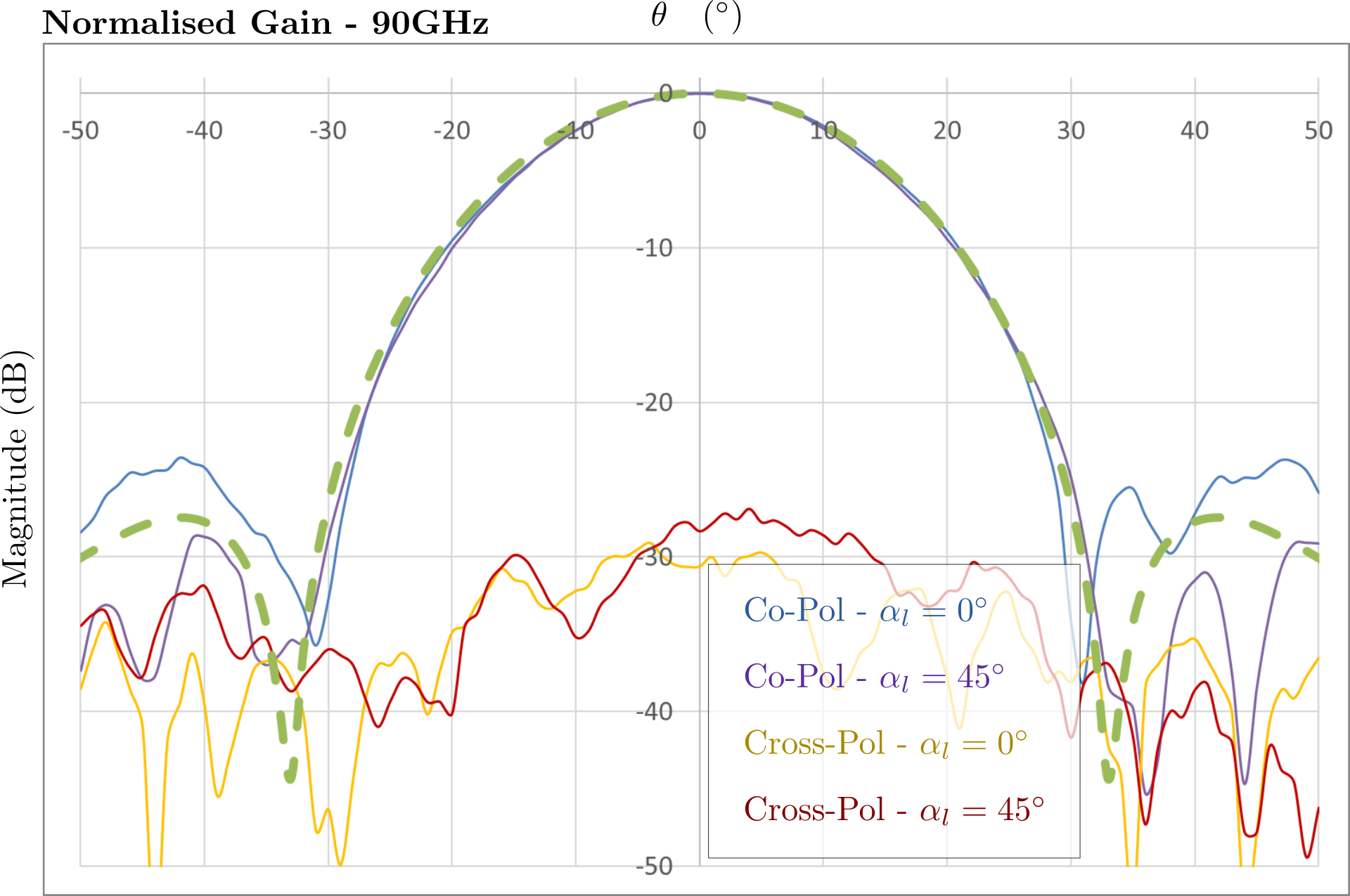}
		\caption{ }
		\label{fig:measres1}
	\end{subfigure}
	\begin{subfigure}[t]{0.45\textwidth}
		\includegraphics[width=\textwidth]{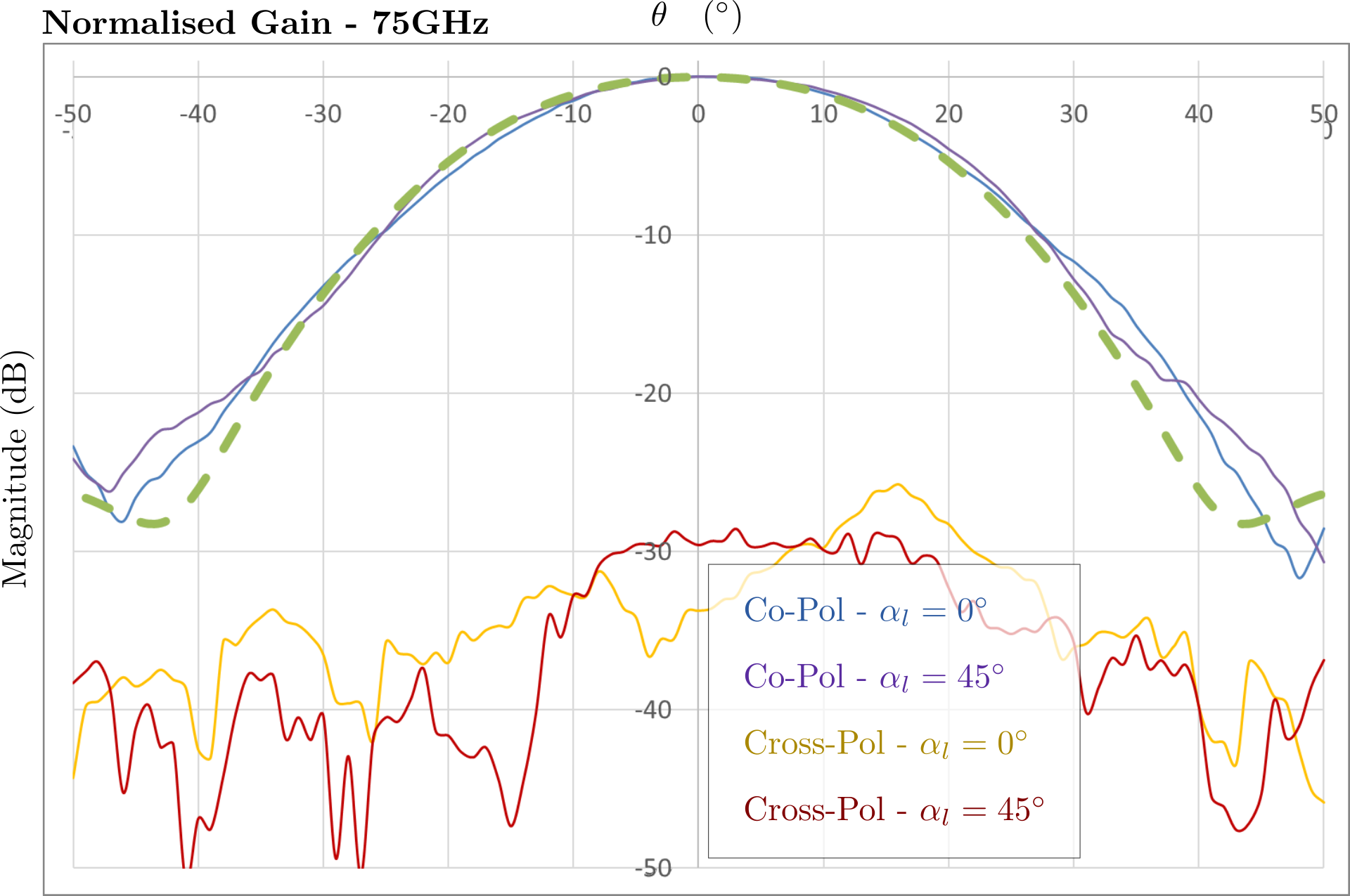}
		\caption{ }
		\label{fig:measres2}
	\end{subfigure}
	\begin{subfigure}[t]{0.45\textwidth}
		\includegraphics[width=\textwidth]{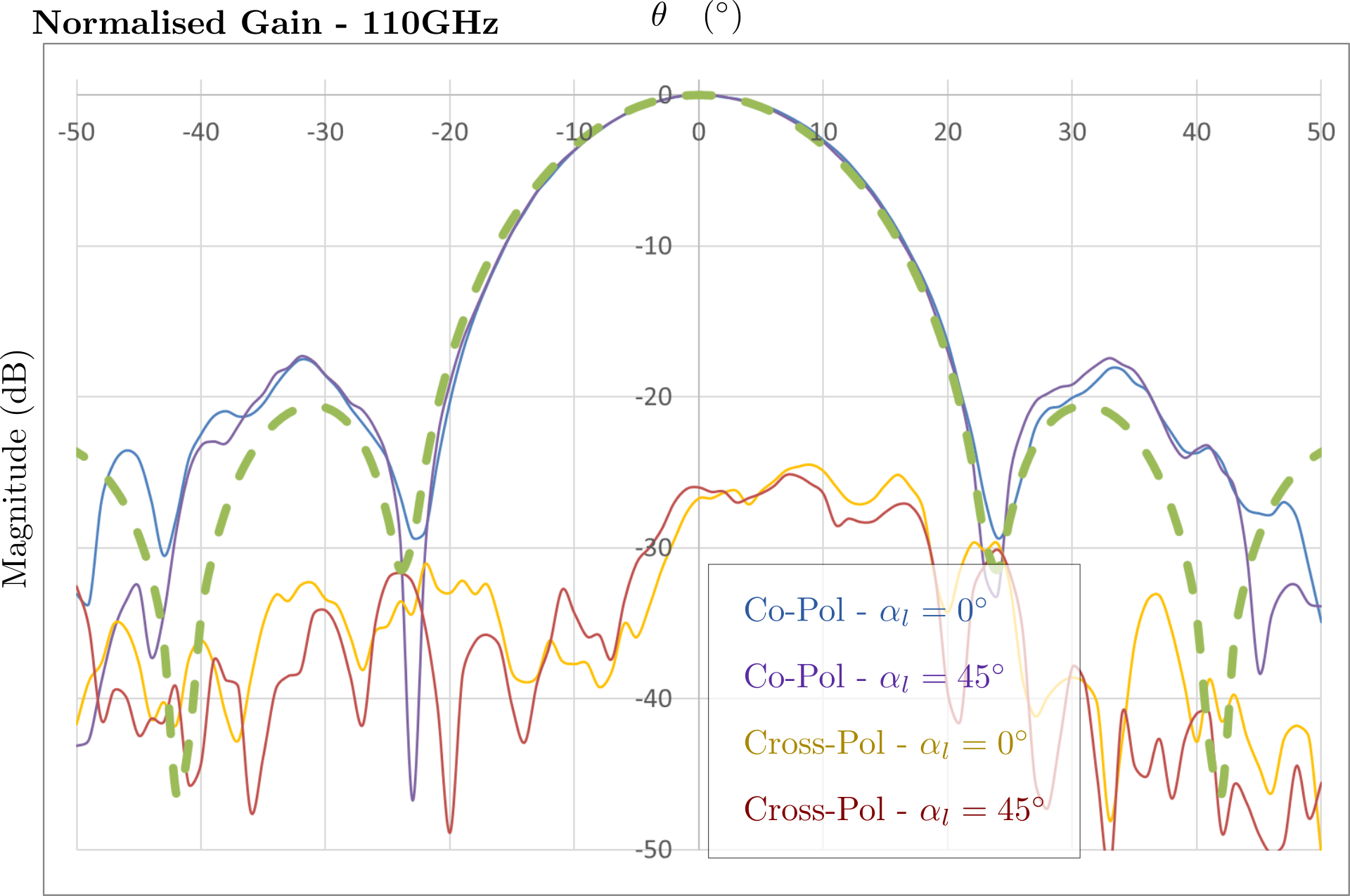}
		\caption{ }
		\label{fig:measres3}
	\end{subfigure}
	\caption{Measured and simulated polarisation leakage and co-polarisation in the H-plane with a lens orientation at $\alpha_l = 0^\circ$ and $\alpha_l = 45^\circ$, at \subref{fig:measres1} $90\si{GHz}$, \subref{fig:measres2} $75\si{GHz}$ and \subref{fig:measres3} $110\si{GHz}$.}
\end{figure}

The mounting structure adopted for this experiment limited the distance between the lenslet and the feed at $Z_0' = 3.32\si{mm}$, still within range of focus $F_0 \pm z_R$. The measured beams at $90\si{GHz}$, $75\si{GHz}$ and $110\si{GHz}$ are presented in Fig. \ref{fig:measres1}, \ref{fig:measres2} and \ref{fig:measres3} respectively, where simulations have been carried out to match the new distance $Z_0'$. For each frequency, the polarisation leakage and the co-polarisation in the H-plane have been measured with the lens orientated at $\alpha_l = 0^\circ$ and $\alpha_l = 45^\circ$. Three main characteristics can be analysed from these data, the main lobe response, the first side-lobe and the polarisation leakage level.

\begin{itemize}
	\item \textbf{Main lobe} The simulated and measured main lobes matches well over the entire bandwidth. Looking at the $90\si{GHz}$ beam, the first null level is reaching $-30\si{dB}$ at \ac{FWE2} equal to $32^\circ$, \ac{FWHM} is $12.1^\circ$, giving a beam width of $2.91mm$. For the two positions of the device, excellent beam homogeneity and gaussianity is observed throughout the full bandwidth.
	\item \textbf{First Side-lobe} The side-lobe peak levels and sizes are in accordance with the simulated response, although there is additional noise contribution from the surrounding at high angle of incidence at which the \ac{RAM} performances decreases. 
	\item \textbf{Polarisation Leakage} The H-plane normalised polarisation leakage levels are consistently below $-25\si{dB}$ over the full bandwidth and for both position of the device, demonstrating expected performances. The peak value for non-normalised data is close to $-80\si{dB}$ and drops down to the $-90\si{dB}$ noise floor of the experimental setup.
\end{itemize}

Whilst these preliminary results are partially validating the lenslet performances, they are incomplete. Co and cross-polarisation measurements in the E and D planes are awaited and polarisation leakage is still to be checked. Despite an experimental setup offering all capabilities to realise such campaign, the mounting strategy presented notable mechanical risks for the lenslet integrity, it also contributed to the far field response at low frequencies and did not enable easy rotation $\alpha_l$ of the lenslet around the \ac{OA}. The choice was made to move onto a different mounting strategy for which a separate measurement campaign will be conducted in the near future.


\section{CONCLUSION}
\label{sec:conclusion}
A comprehensive description of the phase-engineered design procedure of a flat meta-lenslet has been covered, and the resulting performances have been modelled and partially demonstrated experimentally. The lenslet offers a gaussian and homogeneous beam response in the H-plane with sufficient first null level of $30\si{dB}$, an absolute peak gain at $6.65\si{dB}$, or $8.23\si{dBi}$, at $90\si{GHz}$, and polarisation leakage level below $-30\si{dB}$. A complete characterisation of the device will be conducted in the laboratory in the near future, addressing the discrepancy between modelled and measured response at lower frequencies.
A simulated coupling with a planar sinuous antenna is under investigation and will allow a cryogenic testing of both an \ac{OMT} coupled and antenna coupled lenslet to validate proper detection response opening the path for large array fabrication.
While this air-gap lenslet serves as a demonstrator for the design procedure, a fully silicon embedded version is under investigation and should offer a simple solution to \acp{LEKID} coupling with fabrication processes reduced to minimal effort.

\acknowledgments 
This work was supported in part by the Science and Technology Facilities Council [grant number ST/S00033X/1] and in part by NASA under grant/cooperative agreement number NNX17AE85G.

\bibliography{reportbib} 
\bibliographystyle{spiebib} 

\end{document}